\documentclass[sigconf,natbib=true]{acmart}
\AtBeginDocument{%
  \providecommand\BibTeX{{%
    \normalfont B\kern-0.5em{\scshape i\kern-0.25em b}\kern-0.8em\TeX}}}

\setcopyright{acmcopyright}
\copyrightyear{2025}
\acmYear{2025}
\acmDOI{XXXXXXX.XXXXXXX}

\usepackage{balance}

\begin{document}

\title{Interactive Search Intent Prediction with Pre-Search Context}
\title{Interactive Question Prediction using Partial Intent and Pre-Search Context}
\title{Information Need Prediction using Partial Intent and Pre-Search Context}
\title{Interactive Information Need Prediction with Intent and Context}
%\title{Predicting Full Search Intent about a Selected Pre-Search Context while Guided by a Specified Partial Search Intent}

\author{Kevin Ros}
\email{kjros2@illinois.edu}
\affiliation{%
  \institution{University of Illinois Urbana-Champaign}
  \streetaddress{201 N. Goodwin Avenue}
  \city{Urbana}
  \state{Illinois}
  \country{USA}
  \postcode{61801}
}
\author{Dhyey Pandya}
\email{dhyeyhp2@illinois.edu}
\affiliation{%
  \institution{University of Illinois Urbana-Champaign}
  \streetaddress{201 N. Goodwin Avenue}
  \city{Urbana}
  \state{Illinois}
  \country{USA}
  \postcode{61801}
}
\author{ChengXiang Zhai}
\email{czhai@illinois.edu}
\affiliation{%
  \institution{University of Illinois Urbana-Champaign}
  \streetaddress{201 N. Goodwin Avenue}
  \city{Urbana}
  \state{Illinois}
  \country{USA}
  \postcode{61801}
}

\renewcommand{\shortauthors}{TBD, et al.}

\NewDocumentCommand{\kevin}
{ mO{} }{\textcolor{purple}{\textsuperscript{\textit{Kevin}}\textsf{\textbf{\small[#1]}}}}

\NewDocumentCommand{\old}
{ mO{} }{\textcolor{red}{\textsuperscript{\textit{Old}}\textsf{\textbf{\small[#1]}}}}

\NewDocumentCommand{\new}
{ mO{} }{\textcolor{green}{\textsuperscript{\textit{New}}\textsf{\textbf{\small[#1]}}}}

\begin{abstract}
The ability to predict a user's information need would have wide-ranging implications, from saving time and effort to mitigating vocabulary gaps. We study how to interactively predict a user's information need by letting them select a pre-search context (e.g., a paragraph, sentence, or singe word) and specify an optional partial search intent (e.g., "how", "why", "applications", etc.). We examine how various generative language models can explicitly make this prediction by generating a question as well as how retrieval models can implicitly make this prediction by retrieving an answer. We find that this prediction process is possible in many cases and that user-provided partial search intent can help mitigate large pre-search contexts. We conclude that this framework is promising and suitable for real-world applications.

\end{abstract}

\begin{CCSXML}
<ccs2012>
<concept>
<concept_id>10002951.10003317</concept_id>
<concept_desc>Information systems~Information retrieval</concept_desc>
<concept_significance>500</concept_significance>
</concept>
<concept>
<concept_id>10002951.10003317.10003371</concept_id>
<concept_desc>Information systems~Specialized information retrieval</concept_desc>
<concept_significance>500</concept_significance>
</concept>
<concept>
<concept_id>10002951.10003317.10003331.10003271</concept_id>
<concept_desc>Information systems~Personalization</concept_desc>
<concept_significance>300</concept_significance>
</concept>
<concept>
<concept_id>10002951.10003317.10003347.10003348</concept_id>
<concept_desc>Information systems~Question answering</concept_desc>
<concept_significance>300</concept_significance>
</concept>
</ccs2012>
\end{CCSXML}

\ccsdesc[500]{Information systems~Information retrieval}
\ccsdesc[500]{Information systems~Specialized information retrieval}
\ccsdesc[300]{Information systems~Personalization}
\ccsdesc[300]{Information systems~Question answering}

\keywords{information need prediction; pre-search context; search intent; }

%\received{24 January 2024}

\maketitle

\section{Introduction}\label{intro}
% history of ir, query lack of study
In the field of information retrieval, a user's query is considered to be their compromised need - often stemming from a formalized need~\cite{taylor1962process}. Typically, the query offers the only explicit clue of the user's formal information need. Despite the query's prominence in both theoretical and empirical retrieval frameworks, there has been relatively little work on predicting a user's information need before it is manifested in the query itself~\cite{alaofi2022queries}.

% why intent prediction is important
The ability to predict a user's formalized information need is important for a wide variety of reasons. First, such a prediction can help minimize the cognitive overhead of formulating a query, thus helping the user save time and effort. Second, it can help mitigate vocabulary gaps in cases where the user cannot fully articulate their intent through the compromised need (e.g., they are unfamiliar with the search engine or lack background knowledge)~\cite{li2014semantic,van2017remedies}. Finally, the ability to predict the formalized information need can provide perspectives and information beyond immediate needs, thus enabling the user to explore a larger information space. 

Much of the prior work on information need prediction has leveraged a user's pre-search context~\cite{cheng2010actively,liebling2012anticipatory,kong2015predicting}, which has typically been the webpage(s) visited right before a query is entered into a search engine. And it has been found that the query is often \textit{triggered} by such pre-search contexts~\cite{cheng2010actively,liebling2012anticipatory,kong2015predicting}. %For example, the query "applications of information retrieval" may be triggered as a user is reading the Wikipedia page for "Information Retrieval".
But one major limitation here is the rigidity. Different users may be interested in different potions of the webpage, leading to different information needs. Moreover, a user may also want to provide additional keywords to explicitly specify their true intent, which they are unable to do in the prior settings.

Ideally, to enable a user to naturally express the information need with minimum effort, a retrieval system should support interactive interface. It should allow a user to specify the most relevant portion of the pre-search webpage to indicate the context of their information need (e.g., a paragraph, sentence, phrase, or single word), and optionally, allow the user to provide a partial intent \textit{about} the context guiding the predicted need (e.g., "why", "how", "applications", etc.). 
%Using the same example, the user would be able to highlight ``information retrieval" in the Wikipedia page and type in ``applications" as an additional intent-indicating keyword. 
Then, the system could predict the information need by either generating the question or by retrieving documents which answer this question. Besides the benefit of minimizing a user's effort, an interactive interface can be viewed as a step towards providing natural support of human-system collaboration where the user would specify a partial intent, from which the system would then predict the user's full information need. This could be done directly (e.g., thorugh generating a question), or indirectly (e.g., through retrieving an answer). To the best of our knowledge, this is a new challenge that has not been fully studied before. 

%In order to effectively support such an interactive interface, however, we must address a new technical challenge: interactively predicting a user's information need based on both their selected pre-search context and any additional intent-indicating words that they might provide. 

%In this paper, we address this new challenge by studying a novel and general setting of interactive information need prediction,  where the user can select a portion of the webpage to be the pre-search context (i.e., a paragraph, a phrase, a single word), and optionally specify a partial search intent \textit{about} the context (i.e., a shortened query, distilled to just the intent). Following the user's selection of the pre-search context and the optional specification of a partial search intent, we explore the feasibility of information need prediction both directly by generating the question or indirectly by retrieving a paragraph that answers the question.

This setting is illustrated in Figure~\ref{fig:flowchart}, where we show how a user could select the pre-search context from the broader context. In Figure~\ref{fig:flowchart}, the broader context is the Wikipedia page on Information Retrieval and highlighted are three examples of possible user-selected pre-search contexts. For these examples, we show cases where a partial search intent is specified (bottom of figure) and cases where only a pre-search context is selected (right of figure). This described process provides a flexible way for individuals with an information need \textit{about} a certain context to have it satisfied.

In this paper, we explore the technical feasibility of generating questions and retrieving passages for this interactive problem setting, which has not been studied before. We study the following research questions:

\begin{figure*}
    \centering
    \includegraphics[scale=0.35]{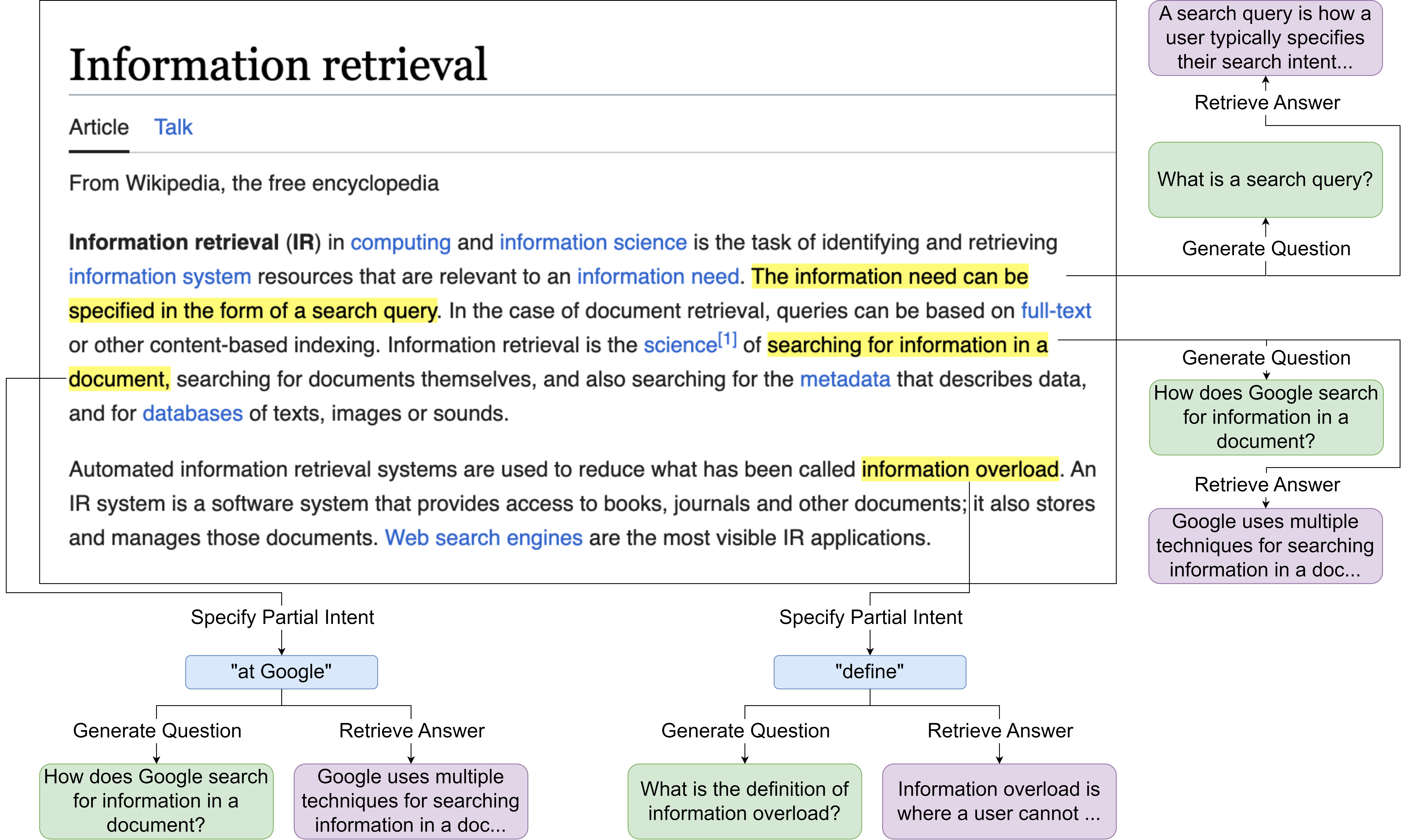}
    \caption{Demonstrating information need prediction via selected pre-search context and specified partial search intent.}
    \label{fig:flowchart}
\end{figure*}

\begin{enumerate}
    \item[\textbf{RQ1}:] How does the amount of selected pre-search context affect the ability of the models to predict the information need?
    \item[\textbf{RQ2}:] How does the amount of specified partial intent affect the ability of the models to predict the information need?
\end{enumerate}
And our contributions are as follows:
\begin{enumerate}
    \item We propose and formalize a novel information need prediction problem, where a user can naturally use an interactive way to select their pre-search context and optionally specify a partial search intent.
    \item We study various state-of-the-art generation and retrieval models for predicting the full need in various settings, and find that the inclusion of a partial intent can help overcome larger pre-search contexts.
    \item We release our code and datasets for future study.\footnote{We will link the code and datasets in the final version of the paper.}
\end{enumerate}

\section{Related Work}

Predicting a user's information need (or more concretely, studying where questions and queries come from) has not received much attention in empirical nor theoretical studies~\cite{alaofi2022queries}. Nevertheless, there have been a handful of early systems proposed to predict search intent, such as Rhodes and Starner's Remembrance Agent~\cite{rhodes1996remembrance} and Budzik and Hammond's Watson~\cite{budzik1999watson}. There have also been empirical studies on the relationships between pre-search context and user queries. Rahurkar and Cucerzan classified cases where a submitted query was related to the most recently browsed news document~\cite{rahurkar2008predicting}. White et al. studied how activity-based context can predict short-term search interest, and found that using context (e.g., navigation behavior, queries) helped with interest prediction~\cite{white2010predicting}.

Cheng et al. proposed that certain queries may be triggered by a browsed webpage, such as queries that result from something being unknown or unclear~\cite{cheng2010actively}. Using query logs, they associated queries with their respective triggering webpages and proposed a ranking and diversification process to suggest likely queries given a webpage. Liebling et al., in one of the first works to explicitly study the problem of performing query-less search to anticipate user needs, compared methods for predicting search intent using the pre-search context, the whole of which they called anticipatory search~\cite{liebling2012anticipatory}. In their work, they took "pre-search context" to be the current URL, and attempted to predict the webpage that a user would try to find next. Finding mixed results after evaluating on log data, Liebling et al. concluded that the URL alone may not be enough to predict the full information need~\cite{liebling2012anticipatory}. Building on this work, Kong et al. examined search logs to find cases where a search query was indeed triggered by the pre-search context, and then proposed a mixture generative model to learn the relationships between the pre-search context and the query~\cite{kong2015predicting}. They tested their models in a query retrieval setting, where the target query was among a candidate pool of 200 queries. However, Kong et al. did not attempt to generate the full information need as all of the candidate queries were assumed to be known. More recently, Ko et al. created a dataset of inquisitive questions, which  consists of questions asked by readers of news articles~\cite{ko2020inquisitive}. Ko et al. also evaluated question generation models across various settings, which is similar to our setting of question generation \textit{without} user-specified partial search intent. They found that the generation task is challenging and has room for improvement. 

%Our studied problem setting is similar to the settings described above, but it differs in multiple key ways.
Compared with this previous related work, our work is novel in that we are the first to study the interplay between user-specified partial search intent and user-selected pre-search context, and how variations of this affect information need prediction. Moreover, we test this explicitly (through question generation) and implicitly (through retrieval). Finally, we do not assume a fixed query pool to predict information needs, and instead focus on the generation of the question itself or the retrieval of the target paragraph from a large collection, thus making our setting more open-ended.

Slightly distinct from our setting is the use of broader contextual features to predict information needs, such as task repetition~\cite{song2016query}, writing tasks~\cite{koskela2018proactive}, or conversations~\cite{lee2018making, torbati2021you}. In general, context has been used to predict search intent, but the vast majority of the work has studied this assuming that a traditional query has already been specified in addition to the context~\cite{merrouni2019toward}, rather than predicting the information need from the context alone. 

Predicting the information need from a partially specified query can be considered query auto-completion (QAC) or query suggestion (QS)~\cite{tahery2020customized}. However, in QAC, the query completions typically match the user's specified query prefix~\cite{tahery2020customized}. In our setting, a user's partially specified search intent need not be a prefix to a predicted full information need. And QS typically aims to present queries that are semantically similar to a previous query on the search result page~\cite{tahery2020customized}, whereas our predictions may not be semantically similar to the specified partial search intent. Distinct from both QAS and QC, we also study the case where a user does not enter any partial search intent, and the full search intent is predicted from the selected context alone.

In conversational question-answering, a user interacts with a retrieval or chat-based system in a conversational style, typically asking follow-up questions in-turn~\cite{choi2018quac,reddy2019coqa}. Our setting is different from conversational question-answering because we focus on information needs that are \textit{triggered} by some context, rather than \textit{answerable} by some context. Moreover, we only study a single interaction, rather than multiple turns (and therefore have no explicit requirement to model conversation history). However, our study could be used to support individual turns in conversational question-answering by formulating queries for or retrieving documents from search systems in RAG-style frameworks~\cite{lewis2020retrieval}.

\section{The Problem Setting}\label{problem setting}
\begin{figure*}
    \centering
    \includegraphics[scale=0.5]{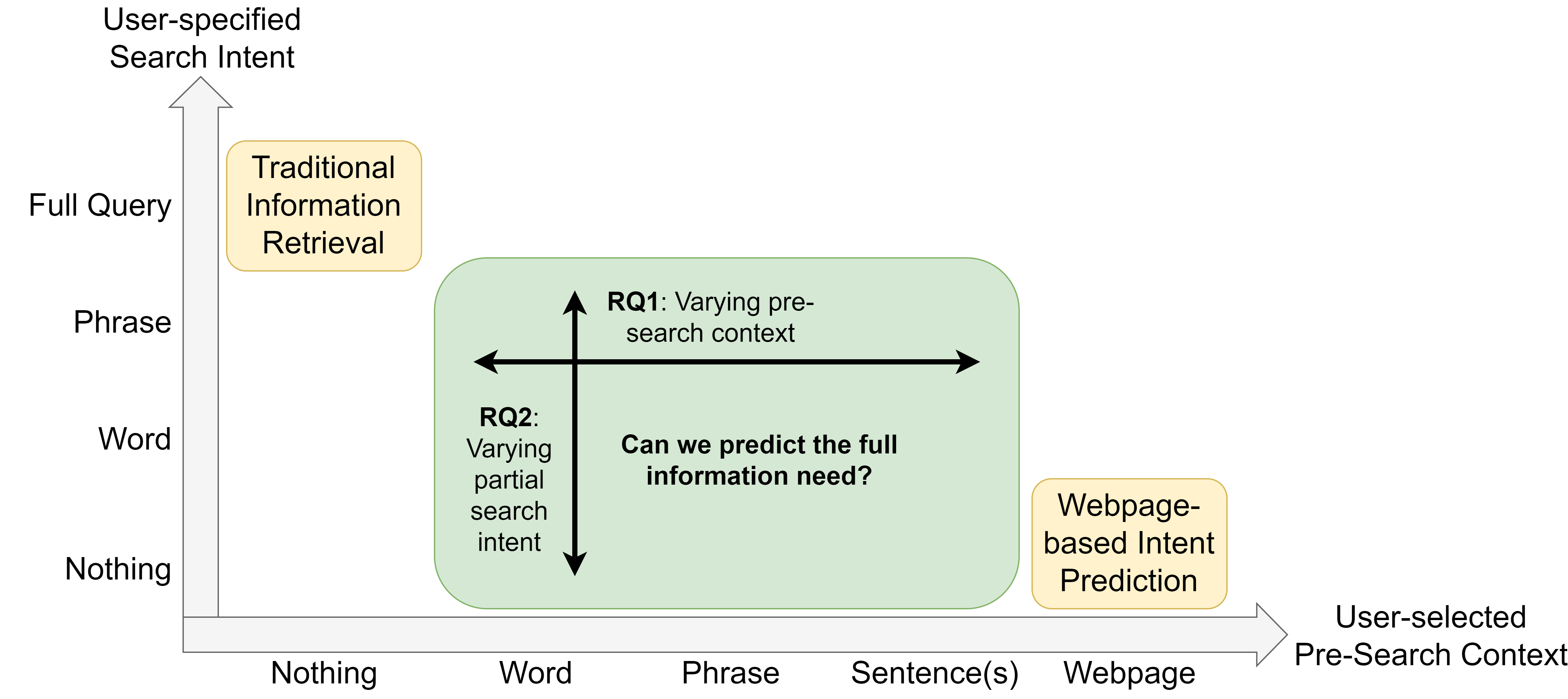}
    \caption{The trade-off between selected pre-search context and specified search intent.}
    \label{fig:intent-context}
\end{figure*}

Motivated by the need to minimize search effort, we study the ability to predict the full information need given varying levels of selected pre-search context and specified partial search intent. Figure~\ref{fig:intent-context} demonstrates the relationship between these two aspects.
The X-axis corresponds to the extent of user-selected pre-search context and the Y-axis corresponds to the extent of the user-specified partial search intent. In the setting of traditional information retrieval, there is no pre-search context, and the user's information need is completely specified by the user typing a full query. Conversely, the prior work that has studied search intent prediction from the full webpage (e.g.,~\cite{cheng2010actively,liebling2012anticipatory,kong2015predicting}) falls into the "Webpage-based Intent Prediction" box, as the entire webpage is considered as the pre-search context and there is no user-specified partial search intent.

We are interested in the middle section of the graph in Figure~\ref{fig:intent-context}, which is highlighted by the green box. This box represents the cases where the user selects the pre-search context from anywhere between a single word to an entire paragraph, and specifies the partial search intent from anywhere between nothing to a phrase. In this box, the information prediction can be thought of as interactive because a user has the ability to select a variable pre-search context and to specify a partial search intent. This is in contrast to traditional information retrieval, where the user cannot select the context, and to webpage-based intent prediction, where the context is fixed to the webpage and there is no specified partial search intent. Studying this problem setting is interesting because it can provide flexibility beyond existing retrieval and recommendation settings, and it can help reduce effort through the broader interactions. Moreover, the answers to our research questions will have direct implications for any applications which aim to implement information need prediction in this way.

\section{Building the Datasets}\label{dataset}
Evaluating the proposed interactive information need prediction task is an open challenge. Ideally, to address this challenge, we would have a dataset of samples consisting of the following fields:
\begin{enumerate}
    \item \textbf{Source}: the broader context of what was being read,
    \item \textbf{Context}: the point of confusion in the source,
    \item \textbf{Intent}: what was asked about the context,
    \item \textbf{Question}: the full, natural language question asked, and
    \item \textbf{Target}: the answer to the question, if applicable.
\end{enumerate}
Because there did not exist any dataset that we could directly use, we augmented two existing datasets to fit our experiments.

The first dataset that we used was the Inquisitive dataset~\cite{ko2020inquisitive}, which is publicly available online~\cite{inquisitive-online}. The inquisitive dataset was built by providing subjects with a sequence of sentences from news articles, and asking them to write down any questions that they had after reading each sentence. The participants were also instructed to select the span of text from the sentence which caused the question. In total, the dataset consists of around 19,000 questions across 1,500 different news articles. We selected the Inquisitive dataset to adapt to our setting because it is one of the closest datasets to our ideal structure, and it contains real questions about content.

In order to adapt the Inquisitive dataset to our setting, we considered each sentence to be the "Source", the corresponding selected span to be the "Context", and the asked question to be the "Question". We augmented the dataset with the "Intent" by using Llama-3-8b-Instruct~\cite{touvron2023llama,llama3-8b} to extract the intent for each question (e..g, to extract items like "who", "why", "how many", "examples", etc.). We optimized our prompt over the first 100 samples of the Inquisitive validation set. Because this dataset did not include answers to the questions asked, we did not add any targets, and only used this dataset in our question generation experiments.

The second dataset that we adapted to our setting was the MS MARCO V2 dataset~\cite{nguyen2016ms}, which is publicly available online~\cite{ms-marco-website}. The MS MARCO dataset is a TREC-style retrieval collection consisting of a large corpus of paragraphs, a set of queries, and a set of relevance judgments for each query over the corpus. We selected the MS MARCO dataset due to its prominence in the field of information retrieval and common structure, so that our techniques here may be applied to other traditional TREC-style datasets. Due to the large size of the dataset, we restricted our analysis to 10,000 queries of the validation set. From this dataset, we considered the query to be the "Question" and its respective positively-judged paragraph to be the "Target". To simulate the "Source" of each query, we performed a keyword-based search of the query over the corpus using BM25~\cite{robertson1994some} via Pyserini~\cite{lin2021pyserini}, and found a set of paragraphs that matched the query in topic but \textit{did not} answer the query. We selected the best-matching paragraph as the "Source". Then, to simulate the "Context" and "Intent", we again relied on a large generative language model, Yi-34B-Chat~\cite{ai2024yi}, which is publicly available on Hugging Face~\cite{yi-34b}. Using this model, we reformulated each query into its "Context" (what the query was about) and its "Intent" (what the query is asking about the context). Following the reformulation, we examined each output and made corrections if necessary.

\begin{table*}[]
    \centering
    \begin{tabular}{|p{3cm}|c|c|c|p{3cm}|}
    \hline
    Source & Context & Intent & Question & Target \\
    \hline
    Defense Minister Pavel Grachev, who has been rumored to be on... & who has been rumored & source & Where do these rumours come from? & N/A \\
    %\hline 
    %The dikes loom high over this Gelderland Province town of 5, 000, virtually ... & town & how many & How many towns are affected? & N/A \\
    \hline
    \hline
    My question isn't about a parrot, it is about a robin egg...  & robin eggs & hatching time & when do robin eggs hatch & ...will hatch 12-14 days after the mother began incubating the eggs \\
    %\hline
    %With metaphors, you don't have to write 'like' or 'as'. For example: His eyes were fireflies... & firefly & light up & how does a firefly light up & this type of light production is called bioluminescence. the method by which fireflies produce light is...\\
    \hline
    \end{tabular}
    \caption{Some samples from the adapted Inquisitive dataset (top) and MS MARCO dataset (bottom).}
    \label{tab:query split examples}
\end{table*}

Table~\ref{tab:query split examples} contains 
a few samples from our adapted datasets. For the MS MARCO samples, we only display the top-retrieved source paragraph. Note that this construction allowed us to create positive and negative retrieval samples. A positive sample is where the inputs (e..g, question, context, intent, and/or source) are mapped to the question's ground-truth target. A negative sample is where the the inputs are mapped to a random target from the corpus. Using these samples, we train models across both the retrieval and generation settings, and we explore how different variations of pre-search context and intent affect model performance.

For the Inquisitive dataset, we followed the training (15,931 samples), validation (1,991 samples), and testing (1,894 samples) splits provided in the original paper~\cite{ko2020inquisitive}. And for the MS MARCO dataset, we randomly assigned each sample into training (80\%, 8,040 samples), validation (10\%, 963 samples), and testing (10\%, 991 samples) sets. For the negative retrieval samples, we selected the target paragraphs from within a query's training/validation/testing split. 

It is important to note that we manually validated the context and intent additions to the MS MARCO dataset but did not validate the intent added to the Inquisitive dataset (outside of the initial validation). This is due to the larger size of the Inquisitive dataset. As a result, our adaption process likely resulted in samples that are not perfect. Therefore, it should be treated as a "silver standard" dataset. Nevertheless, it is interesting to compare the results across datasets, as similar conclusions may imply that human labeling is not strictly necessary for this adaption step.

\section{Experimental Setup}\label{experimental setup}

For our experiments, we explored variations of pre-search context and partial search intent for both generating questions and retrieving target paragraphs. These variations correspond to the area outlined by the green box in Figure~\ref{fig:intent-context}. We varied the partial search intent by either including or excluding the intent extracted from the question, corresponding to the "Type Word/Phrase" or "Type Nothing" locations on the Y-axis in Figure~\ref{fig:intent-context}, respectively. Moreover, we varied the pre-search context by using either the full source or the extracted context, corresponding to the "Select Paragraph" or "Select Word/Phrase" locations on the X-axis in Figure~\ref{fig:intent-context}, respectively. In summary, we tested the generation and retrieval models under the following inputs:
\begin{enumerate}
    \item Question: The baseline run (for retrieval only). Here, the input was the original, unprocessed question. 
    \item Context + Intent: The input was the context and intent from the question reformulation.
    \item Source + Intent: The input was the source and the intent from the question reformulation.
    \item Context: The input was just the context.
    \item Source: The input was just the source.
\end{enumerate}
For answering \textbf{RQ1}, we compared the performance of the models with the inputs of (2) with (3), and (4) with (5). Similarly, for answering \textbf{RQ2}, we compared the performance of the models with the inputs of (2) with (4), and (3) with (5). In the original Inquisitive dataset paper, Ko et al. studied variations of these settings but did not make an explicit distinction for intent, nor did they study question generation from the source sentence alone~\cite{ko2020inquisitive}.

All models were trained and inferenced using a single NVIDIA RTX A5000. For each experiment, the batch size was selected to be as large as possible before causing out-of-memory errors. Moreover, the hyperparameters and their ranges were decided after performing a basic preliminary analysis to achieve a stable training setting. For all statistical significance tests, we used Scipy's standard, independent t-test~\cite{2020SciPy-NMeth}, which is available online~\cite{scipy-website}. This method is commonly accepted in the information retrieval literature~\cite{smucker2007comparison}.

\subsection{Generation}
To explicitly predict the information need by generating the full question, we considered two general approaches: fine-tuning encoder-decoder language models and prompting generative large language models. We considered these two approaches because both are commonly used in today's literature~\cite{liu2023pre}. The models were selected due to their established use as well as their availability online, thus allowing for easier reproducibility. To assess the performance, we measured the ROUGE-\{1,2,L\} and BLEU-\{1,2,3,4\} scores on the validation datasets.  These metrics were chosen due to their complementary nature and common use in evaluating generated text~\cite{celikyilmaz2020evaluation}. We used the ROUGE-\{1,2,L\} measures for testing statistical significance. 

Regarding our fine-tuning approach, we used Flan-T5-Base~\cite{chung2022scaling}, which is a transformer~\cite{vaswani2017attention} encoder-decoder language model based on T5~\cite{raffel2020exploring}. We began with the pre-trained Flan-T5-Base model available online from Hugging Face~\cite{flan-t5-base}. We fine-tuned the model on our training datasets and evaluated the model on our validation sets in each of the variations previously discussed. We used a batch size of 8, 50 warm-up steps, and stopped training when the validation loss increases by more than 5\% of the best-performing model epoch. Additionally, we varied the number of positive samples between \{1,2\} when the model input included a source paragraph, and varied the learning rate over \{1e-4, 5e-5, 1e-5\}. Each training run took approximately two hours. %A full description of the hyperparameters and example training samples are presented in Section~\ref{encoder decoder samples} of the Appendix.

Regarding our prompting approach, we explored the use of both Llama-3-8b-Instruct~\cite{touvron2023llama,llama3-8b} and and Yi-34B-Chat~\cite{ai2024yi,yi-34b}. Both models are decoder-only large transformer models~\cite{radford2019language} fine-tuned for use in dialogue. We quantized~\cite{dettmers2022llm} both models to four bits to perform inference using a single GPU. We used a similar handcrafted prompt construction approach to that described in Section~\ref{dataset} by optimizing the performance on a subset of the validation sets. %The full prompts are presented in Section~\ref{gen llm samples} of the Appendix.

\subsection{Retrieval}
To implicitly predict the information need by retrieving an answering passage, we fine-tuned both bi-encoder and cross-encoder language models. We choose these methods because they are commonly used in the retrieval literature, with bi-encoders often used in the initial retrieval stage~\cite{lin2022pretrained}, and cross-encoders often used in the reranking stage~\cite{reimers-2019-sentence-bert}. We based both approaches on BERT~\cite{devlin2018bert}, an encoder-only pre-trained transformer model. We chose this model due to its generality in various applications. To assess performance, we measured the recall at 10 (R@10) and the mean reciprocal rank (MRR) performance on the validation queries. We selected these two metrics due to their common use in information retrieval and their interpretability for use in web retrieval systems (e.g., such as search engines)~\cite{bama2015survey}. We used the MRR at 10 for testing statistical significance. The validation target paragraphs and the training target paragraphs were used as candidates for retrieval. This amounted to 956 validation samples over 8,873 candidate target paragraphs.

For the bi-encoder model, we used bert-base-uncased, which is publicly available from Hugging Face~\cite{bert-hf}. We trained the model to maximize the cosine similarity between positive training samples and minimize the cosine similarity between negative samples via the cosine embedding loss~\cite{cosineembloss}.
%We used the following method to calculate the %loss~\cite{cosineembloss}:
%\begin{equation}
%    loss(x,y) =  
%    \left\{
%        \begin{array}{lr}
%            1 - cos(x_1, x_2), & \text{if } y = 1\\
%            max(0, cos(x_1, x_2) - margin), & \text{if } y = %-1\\
%        \end{array}
%    \right\}
%\end{equation}
In all of our experiments, we set the margin to 0.5 and set $x_1$ and $x_2$ to be the CLS tokens from the source and target of each sample, respectively. The label indicates if the sample was positive or negative. Moreover, we used a batch size of 16, 50 warm up steps, and stopped training when the validation loss increased by more than 5\% of the best-performing model epoch. We fixed the number of negative samples to 10, varied the number of positive samples (when we had a source paragraph) over \{1,2\}, and varied the learning rate over \{5e-5, 1e-5, 5e-6\}. Each training run took approximately four hours. %The hyperparameters and model input examples are presented in Section~\ref{biencoder samples} of the Appendix.

For the cross-encoder model, we again used Hugging Face's bert-base-uncased but with a classification head. We predicted a single class via mean-squared error, where each positive sample was 1 and each negative sample was 0. We used a batch size of 32, fixed the number of negative samples to 10, varied the number of positive samples (when we had a source paragraph) over \{1,2\}, and varied the learning rate over \{1e-4, 5e-5, 1e-5\}. Each training run took approximately two hours. Note that the retrieval experiments were only performed over the MS MARCO dataset, as the Inquisitive dataset does not have any target passages.

%The full set of hyperparameters and input examples are presented in Section~\ref{crossencoder samples} of the Appendix.

\section{Results}\label{results}
Our main quantitative results are presented in Table~\ref{tab: generation inquisitive}, Table~\ref{tab: generation}, and Table~\ref{tab: retrieval}. In Table~\ref{tab: generation inquisitive} and Table~\ref{tab: generation}, we present the BLEU-\{1,2,3,4\} scores and the ROUGE-\{1,2,L\} scores for the three question generation models on the Inquisitive dataset and the MS MARCO dataset, respectively. Each row corresponds to inputs (2), (3), (4), and (5) discussed at the beginning of Section~\ref{experimental setup}. In all cases, the target output is the generation of the full question. In Table~\ref{tab: retrieval}, we present the R@10 and MRR scores for the two retrieval models over the MS MARCO dataset. Each row corresponds to the inputs (1), (2), (3), (4), and (5) discussed at the beginning of Section~\ref{experimental setup}. In all cases, the target output is the retrieval of the target paragraph. 

We also present qualitative examples in Table~\ref{tab:sample gen queries} and Table~\ref{tab: sample retrieved targets}. In the former, we show the generated questions for the samples in Table~\ref{tab:query split examples} per the different models and varying inputs. In the latter, we show the top-ranked paragraph and target paragraph's rank in the top ten, if present, for the last sample in Table~\ref{tab:query split examples}.

Regarding the overall generative performance presented in both Table~\ref{tab: generation inquisitive} and Table~\ref{tab: generation}, the best-performing model across most settings was the fine-tuned Encoder-Decoder (significant at $p < 0.0249$). The only non-significant differences were the ROUGE-2 measurements in the context to query setting between the Encoder-Decoder and Llama-8b-Instruct for both tables  ($p=0.1333$ and $p=0.0554$, respectively). In fact, the Llama-8b-Instruct model achieved a higher BLEU-1 score than the Encoder-Decoder for Table~\ref{tab: generation}, too. However, we did not find a similar trend for the other BLEU measurements. 

And with the retrieval performance of Table~\ref{tab: retrieval}, the best performing model across all settings was the Cross-Encoder (significant at $p<0.007$). This can be explained by the ability of the Cross-Encoder to attend to both the input and output simultaneously (at a cost of scalability). The performance of the Bi-Encoder, however, was still quite high, and both models were generally competitive with the original query input baseline. In the following subsections, we discuss the results in detail with respect to our research questions.

\begin{table*}
    \centering
    \begin{tabular}{|c|c|c|c|c|c|c|c|c|c|}
    \hline
    Model Name & Input & BLEU-1 & BLEU-2 & BLEU-3 & BLEU-4 & ROUGE-1 & ROUGE-2 & ROUGE-L\\
    \hline
    Encoder-Decoder & Context + Intent & 0.4203 & 0.2597 & 0.1796 & 0.1320 & 0.4409 & 0.2044 & 0.4295\\
    Llama3-8b-Instruct & Context + Intent & 0.3277 & 0.1887 & 0.1231 & 0.0857 & 0.3803 & 0.1559 & 0.3645 \\
    Yi-34B-chat & Context + Intent & 0.3655 & 0.2115 & 0.1393 & 0.0986 & 0.3750 & 0.1502 & 0.36088 \\
    \hline
    Encoder-Decoder & Source + Intent & 0.4016 & 0.2387 & 0.1632 & 0.1200 & 0.3785 & 0.1659 & 0.3675\\
    Llama3-8b-Instruct & Source + Intent & 0.2107 & 0.1036 & 0.0612 & 0.0390 & 0.2637 & 0.1008 & 0.2488\\
    Yi-34B-chat & Source + Intent & 0.2571 & 0.1358 & 0.0856 & 0.0591 & 0.2673 & 0.0984 & 0.2568 \\
    \hline
    Encoder-Decoder & Context & 0.3521 & 0.2030 & 0.1349 & 0.0967 & 0.3301 & 0.1314 & 0.3184\\
    Llama3-8b-Instruct & Context & 0.2822 & 0.1519 & 0.0969 & 0.0672 & 0.3092 & 0.1207 & 0.2979\\
    Yi-34B-chat & Context  & 0.2866 & 0.1409 & 0.0876 & 0.0605 & 0.2625 & 0.0924 & 0.2526\\
    \hline
    Encoder-Decoder & Source & 0.2880 & 0.1393 & 0.0829 & 0.0540 & 0.2175 & 0.0710& 0.2100\\
    Llama3-8b-Instruct & Source & 0.2197 & 0.0868 & 0.0454 & 0.0267 & 0.1730 & 0.0526 & 0.1658\\
    Yi-34B-chat & Source  & 0.2048 & 0.0807 & 0.0444 & 0.0277 & 0.1533 & 0.0421 & 0.1464\\
    \hline
    \end{tabular}
    \caption{Generating the full question using the Inquisitive dataset.}
    \label{tab: generation inquisitive}
\end{table*}

\begin{table*}[]
    \centering
    \begin{tabular}{|c|c|c|c|c|c|c|c|c|c|}
    \hline
    Model Name & Input & BLEU-1 & BLEU-2 & BLEU-3 & BLEU-4 & ROUGE-1 & ROUGE-2 & ROUGE-L\\
    \hline

    %encoderdecoder\_contextintent-to-query\_0001 
    Encoder-Decoder & Context + Intent & 0.7207 & 0.6072 & 0.5098 & 0.4346 & 0.7545 & 0.5311 & 0.7092\\
    %llama\_contextintent-to-query\_5
    %Llama-13b-chat-hf & Context + Intent & 0.4330 & 0.3288 & 0.2558 & 0.2082 & 0.6437 & 0.3760 & 0.5710 \\
    %llama3\_contextintnet-to-query\_2
    Llama3-8b-Instruct & Context + Intent & 0.6593 & 0.4374 & 0.3037 & 0.2187 & 0.6782 & 0.4314 & 0.6310 \\
    % yi\_contextintent-to-query\_4
    Yi-34B-chat & Context + Intent & 0.4624 & 0.3536 & 0.2783 & 0.2238 & 0.6650 & 0.4175 & 0.6063\\
    \hline
    % encoderdecoder\_sourceintent-to-query\_2pos\_0001 
    Encoder-Decoder & Source + Intent & 0.5271 & 0.3915 & 0.2948 &  0.2305 & 0.5414 & 0.3010 & 0.5101\\
    %llama\_sourceintent-to-query\_1
    %Llama-13b-chat-hf & Source + Intent & 0.2423 & 0.1457 & 0.0875 & 0.0550 & 0.3646 & 0.1476 & 0.3235\\
    %llama3\_sourceintnet-to-query\_3
    Llama3-8b-Instruct & Source + Intent & 0.3604 & 0.1608 & 0.0824 & 0.0480 & 0.4186 & 0.1980 & 0.3903 \\
    % yi\_sourceintent-to-query\_3
    Yi-34B-chat & Source + Intent & 0.1577 & 0.0952 & 0.0603 & 0.0404 & 0.3310 & 0.1491 & 0.3068\\
    \hline
     %encoderdecoder\_context-to-query\_00005 
    Encoder-Decoder & Context & 0.5060 & 0.3955 & 0.2989 &  0.2255 & 0.5467 & 0.3274 & 0.5338\\
    %llama_context-to-query_4
    %Llama-13b-chat-hf & Context & 0.1086 & 0.0992 & 0.0909 & 0.0859 & 0.5090 & 0.2855 & 0.5048\\
    %llama3\_context-to-query
    Llama3-8b-Instruct & Context & 0.5853 & 0.3602 & 0.2082 & 0.1993 & 0.5244 & 0.3027 & 0.5053 \\ 
    %yi\_context-to-query\_2
    Yi-34B-chat & Context & 0.1740 & 0.1083 & 0.0673 & 0.0414  & 0.3817 & 0.1920 & 0.3601\\
    \hline
    %encoderdecoder\_source-to-query\_2pos\_0001 
    Encoder-Decoder & Source & 0.3013 & 0.1879 & 0.1158 & 0.0719 & 0.3293 & 0.1343 & 0.3198\\
    %llama\_source-to-query\_2
    %Llama-13b-chat-hf & Source & 0.0994 & 0.0510 & 0.0195 & 0.0075 & 0.2073 & 0.0673 & 0.1986\\
    %llama3\_source-to-query
    Llama3-8b-Instruct & Source & 0.2731 & 0.0990 & 0.0386 & 0.0169 & 0.2740 & 0.1026 & 0.2639 \\
    %yi\_source-to-query\_2
    Yi-34b-chat & Source & 0.0902 & 0.0418 & 0.0219 & 0.0134 & 0.2113 & 0.0734 & 0.2007\\
    \hline
    \end{tabular}
    \caption{Generating the full question using the MS MARCO dataset.}
    \label{tab: generation}
\end{table*}

\begin{table}[]
    \centering
    \begin{tabular}{|c|c|c|c|c|}
    \hline
    Model Name & Input & R@10 & MRR\\
    \hline
    % biencoder\_query-to-target\_1pos\_10neg\_00001
    Bi-Encoder & Question & 0.7490 & 0.5968\\
    %crossencoder\_query-to-targeg\_1pos\_10neg\_00001
    Cross-Encoder & Question & 0.9456 & 0.7490 \\
    \hline
    % biencoder\_contextintent-to-target\_1pos\_10neg\_00001
    Bi-Encoder & Context + Intent  & 0.8285 & 0.6825 \\
    % crossencoder\_contextintent-to-target\_1pos\10neg\_00001
    Cross-Encoder & Context + Intent & 0.9445 & 0.8025  \\
    \hline
    % biencoder\_sourceintent-to-target\_1pos\_10neg\_00001
     Bi-Encoder & Source + Intent & 0.5889 & 0.4181\\
    % crossencoder\_sourceintent-to-target\_ 1pos\_10neg\_00001
     Cross-Encoder & Source + Intent & 0.8400 & 0.5494 \\
    \hline
    %biencoder\_context-to-target\_1pos\_10neg\_00001 
    Bi-Encoder & Context & 0.6967 & 0.5419 \\
    % crossencoder\_context-to-target\_1pos\_10neg\_00001
    Cross-Encoder & Context & 0.8808 & 0.6875 \\
    \hline
    %biencoder\_source-to-target\_1pos\_10neg\_00001 
    Bi-Encoder & Source & 0.5094 & 0.3519\\
    % crossencoder\_source-to-target\_1pos\_10neg\_00005
    Cross-Encoder & Source & 0.7218 & 0.5074 \\
    \hline
    \end{tabular}
    \caption{Retrieving the target passage (MS MARCO).}
    \label{tab: retrieval}
\end{table}

%\begin{table*}[]
%    \centering
%    \begin{tabular}{|c|c|c|}
%    \hline
%     Model & Input & Generated Query\\
%     \hline
%     Encoder-Decoder & Context + Intent & how to light a firefly \\
%     Llama-13b-chat-hf & Context + Intent & firefly light up \\
%     Yi-34B-chat & Context + Intent & How to light up a firefly\\
%     \hline
%     Encoder-Decoder & Source + Intent & fireflies light up eyes \\
%      Llama-13b-chat-hf & Source + Intent & light up like fireflies \\
%     Yi-34B-chat & Source + Intent & How to use metaphors without 'like' or 'as' \\
%     \hline
%     Encoder-Decoder & Context & what is a firefly \\
%     Llama-13b-chat-hf & Context & firefly \\
%     Yi-34B-chat & Context & Firefly context: What is the significance of fireflies in different cultures? \\
%     \hline
%     Encoder-Decoder & Source & what is a metaphor \\
%     Llama-13b-chat-hf & Source &  metaphors examples\\
%     Yi-34B-chat & Source &  How to use metaphors effectively in writing\\
%     \hline
%    \end{tabular}
%    \caption{The generated queries for the query "how does a firefly light up". The inputs for this query are specified in %Table~\ref{tab:query split examples}.}
%    \label{tab:sample gen queries}
%\end{table*}

\begin{table*}[]
    \centering
    \begin{tabular}{|c|c|c|}
    \hline
     Model & Input & Generated Question\\
     \hline
     Encoder-Decoder & Context + Intent & Who is the source of this rumor?\\
     Llama3-8b-Instruct & Context + Intent & who is the source of the rumors about who has been rumored? \\
     Yi-34B-chat & Context + Intent & Who has been rumored as the source of the information?\\
     \hline
     Encoder-Decoder & Source + Intent & What is the source of this information? \\
     Llama3-8b-Instruct & Source + Intent &  what is the source of the news about defense minister pavel grachev being hospitalized?\\
     Yi-34B-chat & Source + Intent &   What is the name of the Defense Minister who has been rumored to be on his way out ...\\
     \hline
     Encoder-Decoder & Context & Who is rumored to have been the sad guy?\\
     Llama3-8b-Instruct & Context & who has been rumored? \\
     Yi-34B-chat & Context & Who has been rumored?\\
     \hline
     Encoder-Decoder & Source & Why is he rumored to be on his way out? \\
     Llama3-8b-Instruct & Source & what is the chechen crisis? \\
     Yi-34B-chat & Source &  What is the Chechen crisis?\\
     \hline
     \hline
     Encoder-Decoder & Context + Intent & how long do robin eggs take to hatch\\
     Llama3-8b-Instruct & Context + Intent & what is the hatching time for robin eggs? \\
     Yi-34B-chat & Context + Intent & What is the hatching time for robin eggs?\\
     \hline
     Encoder-Decoder & Source + Intent & how long does a robin egg take to hatch\\
     Llama3-8b-Instruct & Source + Intent & what is the hatching time for a robin egg \\
     Yi-34B-chat & Source + Intent &  How long can a robin egg survive without incubation before it loses its ability to hatch?\\
     \hline
     Encoder-Decoder & Context & how many eggs do robin eggs hatch\\
     Llama3-8b-Instruct & Context & what are robin eggs \\
     Yi-34B-chat & Context & What are robin eggs and why are they blue?\\
     \hline
     Encoder-Decoder & Source & what is the robin egg \\
     Llama3-8b-Instruct & Source &  what to do with a robin egg\\
     Yi-34B-chat & Source & How long can a robin egg survive without incubation? \\
     \hline
    \end{tabular}
    \caption{Examples of generated questions from the Inquisitive dataset (top) and MS MARCO dataset (bottom).}
    \label{tab:sample gen queries}
\end{table*}

\begin{table*}[]
    \centering
    \begin{tabular}{|c|c|c|p{9cm}|}
    \hline
    Model & Input & Target Rank & Top-Retrieved Target Paragraph \\
    \hline
    Bi-Encoder & Query & Not in Top 10 & fire stone can be obtained through bill's grandfather on route 25...\\
    \hline
     Cross-Encoder & Query & 1 & this type of light production is called bioluminescence. the method... \\
    \hline
    \hline
    Bi-Encoder & Context + Intent & 5 & fire stone can be obtained through bill's grandfather on route 25...\\
    \hline
    Cross-Encoder & Context + Intent & 1 & this type of light production is called bioluminescence. the method ... \\
    \hline
    \hline
    Bi-Encoder & Source + Intent & Not in Top 10 & some of the light that falls on a water drop enters the drop. as it...\\
    \hline
    Cross-Encoder & Source + Intent & 2 & while simple metaphors make a direct comparison between two ... \\
    \hline
    \hline
    Bi-Encoder & Context & Not in Top 10 &  fire stone can be obtained through bill's grandfather on route 25...\\
    \hline
    Cross-Encoder & Context & 1 & this type of light production is called bioluminescence. the method...\\
    \hline
    \hline
    Bi-Encoder & Source  & Not in Top 10 & ...brightness is an attribute of visual perception in which a source... \\
    \hline
    Cross-Encoder & Source  & 1 & this type of light production is called bioluminescence. the method... \\
    \hline
    \end{tabular}
    \caption{The top-ranked retrieved target paragraphs for the query "how does a firefly light up".}
    \label{tab: sample retrieved targets}
\end{table*}

\subsection{RQ1: Varying the pre-search context}
For \textbf{RQ1}, we were interested in examining how changing the amount of user-selected pre-search context affected the ability of the models to predict the full information need. From the generation results in Table~\ref{tab: generation inquisitive}, we found that both the BLEU and ROUGE scores are generally lower for when the source paragraph is used compared to when the question context is used (significant at $p<4.2016e-6$), and we found a similar trend for Table~\ref{tab: generation} (significant at $p<1.1144e-55$). This can be explained by the possibility of the sources introducing additional noise into the prediction process. 

%The decrease in performance was also more evident with respect to the BLEU scores than with respect to the ROUGE scores, indicating that the additional context affected the precision of the generative models more than the recall.

For the retrieval results presented in Table~\ref{tab: retrieval}, we saw a similar decrease in using the source compared to just the question context across both models (significant at $p < 7e-7$). However, the decrease here was not as evident as the generation results. This may be due to both the R@10 and the MRR being more forgiving measures (i.e., small ranking differences do not affect the score as much  as small generation changes).

In Table~\ref{tab:sample gen queries}, we present the generated questions in all cases for two samples of Table~\ref{tab:query split examples}: "Where do these rumours come from?" from the Inquisitive dataset and "when do robin eggs hatch" from the MS MARCO dataset. The addition of a larger pre-search context tended to result in the models generating questions not necessarily about the desired context. 

In Table~\ref{tab: sample retrieved targets}, we present the the top-retrieved paragraph and the top-ten rank of the target paragraph, if present, for the question "how do fireflies light up". The Cross-Encoder provides an interesting example, as in all settings the target paragraph was present in the top-ten. As expected, the introduction of a larger context with intent ("Source + Intent") slightly degraded the model's performance. However, in the case of just the larger context ("Source"), the model retrieved the correct result. And even though there was a slight degradation of performance, the top-retrieved paragraphs were still relevant to the source paragraph (e.g., the Cross-Encoder's "Source + Intent", discussing metaphors), implying that there would still be potential utility in these cases. In the case of "Source", the metaphor target paragraph was still included in the top-ten retrieved paragraphs. Comparing the cases for using the source paragraph to using the question context, we found that the use of the source paragraph caused similar, albeit smaller deviations to the models. 

In conclusion, we found that the amount of pre-search context does play a role on the ability of the models to predict the full information need, as the larger pre-search context can cause deviations from the the target information need to other concepts present in the source. This suggests that it would be beneficial to allow and encourage a user to highlight the most relevant part of the source in the interactive formulation. Moreover, retrieval appears to perform more consistently than generation, albeit at the cost of interpretability. This might be mitigated by the generation of multiple questions to maximize coverage.

\subsection{RQ2: Varying the partial search intent}

For \textbf{RQ2}, we were interested in seeing how the presence of the user-specified partial search intent affected the ability of the models to predict the full information need. Beginning with the Inquisitive generation results of Table~\ref{tab: generation inquisitive}, we found that including the partial search intent improved generation performance (significant at $p < 0.0018$). For MS MARCO generation results presented in Table~\ref{tab: generation}, we found  a similar improvement (significant at $p<2.558e-17$). Interestingly, the models performed similarly when the input is the "Source + Intent" and when the input is the "Context" (more evident in Table~\ref{tab: generation}) - implying that specifying the partial intent is enough for the model to anticipate the specific context from the general source. This can also be seen in Table~\ref{tab:sample gen queries}. For example, in the case of the Encoder-Decoder model on MS MARCO, the generated question from the source paragraph alone is about the number of robin eggs. But the inclusion of the intent ("hatch time"), guided the model to asking about the egg's hatch time.

For the retrieval models of Table~\ref{tab: retrieval}, we found that the intent played a similar role. The intent helped in the "Context + Intent" input setting compared to just the "Context", and we saw a similar improvement with "Source + Intent" over just "Source" (significant at $p < 0.0386$, except for the Cross-Encoder "Source + Intent" and "Source", with $p=0.6748$). In other words, it was beneficial to add intent words to a general source during interactive information need prediction, which can help guide the model towards selecting the correct, specific context when retrieving target paragraphs.

However, in Table~\ref{tab: sample retrieved targets}, we found that the Bi-Encoder retrieval model was unable to retrieve the target paragraph in the top-ten, except for the "Context + Intent" setting. Here, the presence of the intent did not seem to help guide the model with respect to the top-ranked paragraph. The Cross-Encoder model performed much better for this question: retrieving the top paragraph across all four settings, and independent of intent. This implies that, for some cases, the Cross-Encoder may be predicting the intent from the source paragraph or context itself. In conclusion, we found that a partially-specified search intent positively affects generation and retrieval performance, but there were cases where the intent was seemingly ignored by the models.

\section{Discussion}
The main takeaway from this study is as follows. a large amounts of pre-search context can distract both the generation and retrieval models, but minimal specified partial search intent can help mitigate these distractions, suggesting that allowing a user to add partial intent words to a pre-search context is an effective way to enable interactive information need prediction with minimal user effort.  

In essence, the interactive nature of our framework can be viewed as a "conversational query formulation" strategy - allowing users to resolve information needs in a more conversational way. Naturally, there is a trade-off between the amount of effort required by the user to specify their need and the ability for the system to formulate a precise resolution. This is why flexibility is key. Providing the user with the option to select a pre-search context and optionally specify a partial intent can help cover cases where minimal effort is required (e.g., only a highlighted word or phrase) and where maximal effort is required (e.g., having to type a full question), as we demonstrate in Figure~\ref{fig:flowchart} and Figure~\ref{fig:intent-context}. Such a trade-off may be resolved through multi-turn interactions, but we leave this to future work. 

There are a few limitations of this study that are worth noting. First, the datasets used are limited in the sense that they were not originally constructed for this purpose. Despite our efforts towards adapting existing datasets and finding similar results across both, future work could look to improve the quality of adapted datasets for this setting, such as through more robust processing. Or, collecting and building datasets specifically for this purpose will help the research community benchmark and build on results in a more controlled setting. Towards this direction, a more user-centric approach, e.g., through user studies, may reveal variations in question preference across different backgrounds, mediums, or goals. Such user studies may also lead to applications, which would allow us to validate this framework in real-world use cases. One example of an application could look like a browser-based extension that allows a user to highlight any amount of text and then presents the user with a list of questions that were predicted from the selected context. If the user is not satisfied with the questions, then they can type a partial search intent and re-generate questions or retrieve webpages which answer the predicted information need. Such an application would allow us to study the ways in which users prefer to interact with search intent prediction systems, and it would have the potential to help users reduce their effort through the ability to avoid specifying the full question every time.

\section{Conclusion}

In this paper, we proposed a novel and flexible framework that enables information need prediction through the specification of a pre-search context and an optional partial intent, and we studied how to predict this information need in various application contexts. To do this, we modified two existing datasets, the Inquisitive question dataset and the MS MARCO dataset, and then fine-tuned and prompted various generation and retrieval models across settings to generate questions or retrieve answers, respectively. We found that the increase of selected pre-search context can negatively affect performance, but the increase of specified partial search intent can positively affect performance, possibly mitigating larger contexts. Overall, our study has shown that supporting a user to predict their information need with a combination of pre-search context selection and partial intent specification is a promising novel interaction strategy.

%First, the MS MARCO dataset (and other TREC-style collections) has many more queries which can be used to construct a larger simulated dataset. Training (perhaps larger or more powerful) generative and retrieval models on a dataset of increased size should improve the general performance. Second, there are other variations of input and output in our proposed problem setting that went beyond the scope of this paper, but would be interesting for future work. For example, predicting multiple specific contexts from a larger source paragraph (such as keywords or key sentences) to help guide downstream retrieval and generation across multiple possible queries may be a way to mitigate the noise present in larger contexts. Third, the preferences of users who search in the proposed interactive way are not known, as there are not many user studies on search intent prediction. Studying these preferences may lead to new technique or system designs that can leverage the proposed methodology to reduce user effort, or even help completely mitigate the need for a query in certain contextual situations. 

\section{Ethical Considerations}
As with any study on a new technique, there is always a potential for negative social impacts. Fundamentally, this research deals with the prediction of information needs via text generation and retrieval. Beyond the traditional risks of presenting harmful content, if implemented, it is unclear how this prediction process would affect the information seeking behavior of users. For example, the set of presented questions may influence the goals of its users, which could be innocuous or malicious. Should this research be implemented in an application-based setting, designers should take care to ensure that the prediction process is as beneficial as possible.

\begin{comment}
\section{Limitations}
There are a few limitations of our proposed study. First, our dataset size only covered a small subset of the entire MS MARCO dataset. Although our conclusions were promising for applications towards predicting search intent, increasing the dataset size would help make these conclusions much stronger. Moreover, incorporating real user feedback in the generation and retrieval process, such as relevance judgments or labels, would also help ground these conclusions for real-world applications.

Second, our findings rely on assumptions made during the dataset construction process which may not hold in certain cases. For example, certain retrieved source paragraphs may match a query's context as a homonym, rather than the indented context use. More generally, there is no guarantee that a user would ask the queries after reading one of the source paragraphs. This can be mitigated in the future by more expansive labeling during the dataset construction process, or by designing dataset collection studies specific to in-context search intent prediction.

Third, due to the limited computing %compute and cost 
resources, we could not explore the use of larger language models (either via fine-tuning or by prompting) in our experiments. Our findings are meant to act as baselines: demonstrating what is possible with respect to search intent prediction. We have no doubt that increasing model sizes or dedicating time to fine-tuning larger language models will improve performance.

\end{comment}
\balance

\bibliography{sample-base}


\begin{thebibliography}{47}

%%% the Web address.

\ifx \showCODEN    \undefined \def \showCODEN     #1{\unskip}     \fi
\ifx \showDOI      \undefined \def \showDOI       #1{#1}\fi
\ifx \showISBNx    \undefined \def \showISBNx     #1{\unskip}     \fi
\ifx \showISBNxiii \undefined \def \showISBNxiii  #1{\unskip}     \fi
\ifx \showISSN     \undefined \def \showISSN      #1{\unskip}     \fi
\ifx \showLCCN     \undefined \def \showLCCN      #1{\unskip}     \fi
\ifx \shownote     \undefined \def \shownote      #1{#1}          \fi
\ifx \showarticletitle \undefined \def \showarticletitle #1{#1}   \fi
\ifx \showURL      \undefined \def \showURL       {\relax}        \fi
% invisible to TeX
\providecommand\bibfield[2]{#2}
\providecommand\bibinfo[2]{#2}
\providecommand\natexlab[1]{#1}
\providecommand\showeprint[2][]{arXiv:#2}

\bibitem[AI et~al\mbox{.}(2024)]%
        {ai2024yi}
\bibfield{author}{\bibinfo{person}{01. AI}, \bibinfo{person}{:}, \bibinfo{person}{Alex Young}, \bibinfo{person}{Bei Chen}, \bibinfo{person}{Chao Li}, \bibinfo{person}{Chengen Huang}, \bibinfo{person}{Ge Zhang}, \bibinfo{person}{Guanwei Zhang}, \bibinfo{person}{Heng Li}, \bibinfo{person}{Jiangcheng Zhu}, \bibinfo{person}{Jianqun Chen}, \bibinfo{person}{Jing Chang}, \bibinfo{person}{Kaidong Yu}, \bibinfo{person}{Peng Liu}, \bibinfo{person}{Qiang Liu}, \bibinfo{person}{Shawn Yue}, \bibinfo{person}{Senbin Yang}, \bibinfo{person}{Shiming Yang}, \bibinfo{person}{Tao Yu}, \bibinfo{person}{Wen Xie}, \bibinfo{person}{Wenhao Huang}, \bibinfo{person}{Xiaohui Hu}, \bibinfo{person}{Xiaoyi Ren}, \bibinfo{person}{Xinyao Niu}, \bibinfo{person}{Pengcheng Nie}, \bibinfo{person}{Yuchi Xu}, \bibinfo{person}{Yudong Liu}, \bibinfo{person}{Yue Wang}, \bibinfo{person}{Yuxuan Cai}, \bibinfo{person}{Zhenyu Gu}, \bibinfo{person}{Zhiyuan Liu}, {and} \bibinfo{person}{Zonghong Dai}.} \bibinfo{year}{2024}\natexlab{}.
\newblock \bibinfo{title}{Yi: Open Foundation Models by 01.AI}.
\newblock
\newblock
\showeprint[arxiv]{2403.04652}~[cs.CL]


\bibitem[Alaofi et~al\mbox{.}(2022)]%
        {alaofi2022queries}
\bibfield{author}{\bibinfo{person}{Marwah Alaofi}, \bibinfo{person}{Luke Gallagher}, \bibinfo{person}{Dana McKay}, \bibinfo{person}{Lauren~L Saling}, \bibinfo{person}{Mark Sanderson}, \bibinfo{person}{Falk Scholer}, \bibinfo{person}{Damiano Spina}, {and} \bibinfo{person}{Ryen~W White}.} \bibinfo{year}{2022}\natexlab{}.
\newblock \showarticletitle{Where Do Queries Come From?}. In \bibinfo{booktitle}{\emph{Proceedings of the 45th International ACM SIGIR Conference on Research and Development in Information Retrieval}}. \bibinfo{pages}{2850--2862}.
\newblock


\bibitem[Bama et~al\mbox{.}(2015)]%
        {bama2015survey}
\bibfield{author}{\bibinfo{person}{S~Sathya Bama}, \bibinfo{person}{MI Ahmed}, {and} \bibinfo{person}{A Saravanan}.} \bibinfo{year}{2015}\natexlab{}.
\newblock \showarticletitle{A survey on performance evaluation measures for information retrieval system}.
\newblock \bibinfo{journal}{\emph{International Research Journal of Engineering and Technology}} \bibinfo{volume}{2}, \bibinfo{number}{2} (\bibinfo{year}{2015}), \bibinfo{pages}{1015--1020}.
\newblock


\bibitem[Budzik and Hammond(1999)]%
        {budzik1999watson}
\bibfield{author}{\bibinfo{person}{Jay Budzik} {and} \bibinfo{person}{Kristian Hammond}.} \bibinfo{year}{1999}\natexlab{}.
\newblock \showarticletitle{Watson: Anticipating and contextualizing information needs}. In \bibinfo{booktitle}{\emph{Proceedings of the ASIST Annual Meeting}}, Vol.~\bibinfo{volume}{36}. \bibinfo{pages}{727--40}.
\newblock


\bibitem[Celikyilmaz et~al\mbox{.}(2020)]%
        {celikyilmaz2020evaluation}
\bibfield{author}{\bibinfo{person}{Asli Celikyilmaz}, \bibinfo{person}{Elizabeth Clark}, {and} \bibinfo{person}{Jianfeng Gao}.} \bibinfo{year}{2020}\natexlab{}.
\newblock \showarticletitle{Evaluation of text generation: A survey}.
\newblock \bibinfo{journal}{\emph{arXiv preprint arXiv:2006.14799}} (\bibinfo{year}{2020}).
\newblock


\bibitem[Cheng et~al\mbox{.}(2010)]%
        {cheng2010actively}
\bibfield{author}{\bibinfo{person}{Zhicong Cheng}, \bibinfo{person}{Bin Gao}, {and} \bibinfo{person}{Tie-Yan Liu}.} \bibinfo{year}{2010}\natexlab{}.
\newblock \showarticletitle{Actively predicting diverse search intent from user browsing behaviors}. In \bibinfo{booktitle}{\emph{Proceedings of the 19th international conference on World wide web}}. \bibinfo{pages}{221--230}.
\newblock


\bibitem[Choi et~al\mbox{.}(2018)]%
        {choi2018quac}
\bibfield{author}{\bibinfo{person}{Eunsol Choi}, \bibinfo{person}{He He}, \bibinfo{person}{Mohit Iyyer}, \bibinfo{person}{Mark Yatskar}, \bibinfo{person}{Wen-tau Yih}, \bibinfo{person}{Yejin Choi}, \bibinfo{person}{Percy Liang}, {and} \bibinfo{person}{Luke Zettlemoyer}.} \bibinfo{year}{2018}\natexlab{}.
\newblock \showarticletitle{QuAC: Question answering in context}.
\newblock \bibinfo{journal}{\emph{arXiv preprint arXiv:1808.07036}} (\bibinfo{year}{2018}).
\newblock


\bibitem[Chung et~al\mbox{.}(2022)]%
        {chung2022scaling}
\bibfield{author}{\bibinfo{person}{Hyung~Won Chung}, \bibinfo{person}{Le Hou}, \bibinfo{person}{Shayne Longpre}, \bibinfo{person}{Barret Zoph}, \bibinfo{person}{Yi Tay}, \bibinfo{person}{William Fedus}, \bibinfo{person}{Eric Li}, \bibinfo{person}{Xuezhi Wang}, \bibinfo{person}{Mostafa Dehghani}, \bibinfo{person}{Siddhartha Brahma}, \bibinfo{person}{Albert Webson}, \bibinfo{person}{Shixiang~Shane Gu}, \bibinfo{person}{Zhuyun Dai}, \bibinfo{person}{Mirac Suzgun}, \bibinfo{person}{Xinyun Chen}, \bibinfo{person}{Aakanksha Chowdhery}, \bibinfo{person}{Sharan Narang}, \bibinfo{person}{Gaurav Mishra}, \bibinfo{person}{Adams Yu}, \bibinfo{person}{Vincent Zhao}, \bibinfo{person}{Yanping Huang}, \bibinfo{person}{Andrew Dai}, \bibinfo{person}{Hongkun Yu}, \bibinfo{person}{Slav Petrov}, \bibinfo{person}{Ed~H. Chi}, \bibinfo{person}{Jeff Dean}, \bibinfo{person}{Jacob Devlin}, \bibinfo{person}{Adam Roberts}, \bibinfo{person}{Denny Zhou}, \bibinfo{person}{Quoc~V. Le}, {and} \bibinfo{person}{Jason Wei}.}
  \bibinfo{year}{2022}\natexlab{}.
\newblock \bibinfo{title}{Scaling Instruction-Finetuned Language Models}.
\newblock
\newblock
\urldef\tempurl%
\url{https://doi.org/10.48550/ARXIV.2210.11416}
\showDOI{\tempurl}


\bibitem[community(2024)]%
        {scipy-website}
\bibfield{author}{\bibinfo{person}{The~SciPy community}.} \bibinfo{year}{2024}\natexlab{}.
\newblock \bibinfo{booktitle}{\emph{scipy.stats.ttest\_ind}}.
\newblock
\urldef\tempurl%
\url{https://docs.scipy.org/doc/scipy/reference/generated/scipy.stats.ttest_ind.html}
\showURL{%
\tempurl}


\bibitem[Dettmers et~al\mbox{.}(2022)]%
        {dettmers2022llm}
\bibfield{author}{\bibinfo{person}{Tim Dettmers}, \bibinfo{person}{Mike Lewis}, \bibinfo{person}{Younes Belkada}, {and} \bibinfo{person}{Luke Zettlemoyer}.} \bibinfo{year}{2022}\natexlab{}.
\newblock \showarticletitle{Llm. int8 (): 8-bit matrix multiplication for transformers at scale}.
\newblock \bibinfo{journal}{\emph{arXiv preprint arXiv:2208.07339}} (\bibinfo{year}{2022}).
\newblock


\bibitem[Devlin et~al\mbox{.}(2018)]%
        {devlin2018bert}
\bibfield{author}{\bibinfo{person}{Jacob Devlin}, \bibinfo{person}{Ming-Wei Chang}, \bibinfo{person}{Kenton Lee}, {and} \bibinfo{person}{Kristina Toutanova}.} \bibinfo{year}{2018}\natexlab{}.
\newblock \showarticletitle{Bert: Pre-training of deep bidirectional transformers for language understanding}.
\newblock \bibinfo{journal}{\emph{arXiv preprint arXiv:1810.04805}} (\bibinfo{year}{2018}).
\newblock


\bibitem[Face(2024a)]%
        {yi-34b}
\bibfield{author}{\bibinfo{person}{Hugging Face}.} \bibinfo{year}{2024}\natexlab{a}.
\newblock \bibinfo{booktitle}{\emph{01-ai/Yi-34B-Chat - Hugging Face}}.
\newblock
\urldef\tempurl%
\url{https://huggingface.co/01-ai/Yi-34B-Chat}
\showURL{%
\tempurl}


\bibitem[Face(2024b)]%
        {bert-hf}
\bibfield{author}{\bibinfo{person}{Hugging Face}.} \bibinfo{year}{2024}\natexlab{b}.
\newblock \bibinfo{booktitle}{\emph{google-bert/bert-base-uncased - Hugging Face}}.
\newblock
\urldef\tempurl%
\url{https://huggingface.co/google-bert/bert-base-uncased}
\showURL{%
\tempurl}


\bibitem[Face(2024c)]%
        {flan-t5-base}
\bibfield{author}{\bibinfo{person}{Hugging Face}.} \bibinfo{year}{2024}\natexlab{c}.
\newblock \bibinfo{booktitle}{\emph{google/flan-t5-base - Hugging Face}}.
\newblock
\urldef\tempurl%
\url{https://huggingface.co/google/flan-t5-base}
\showURL{%
\tempurl}


\bibitem[Face(2024d)]%
        {llama3-8b}
\bibfield{author}{\bibinfo{person}{Hugging Face}.} \bibinfo{year}{2024}\natexlab{d}.
\newblock \bibinfo{booktitle}{\emph{meta-llama/Meta-Llama-3-8B-Instruct - Hugging Face}}.
\newblock
\urldef\tempurl%
\url{https://huggingface.co/google/flan-t5-base}
\showURL{%
\tempurl}


\bibitem[Ko et~al\mbox{.}(2020a)]%
        {ko2020inquisitive}
\bibfield{author}{\bibinfo{person}{Wei-Jen Ko}, \bibinfo{person}{Te-yuan Chen}, \bibinfo{person}{Yiyan Huang}, \bibinfo{person}{Greg Durrett}, {and} \bibinfo{person}{Junyi~Jessy Li}.} \bibinfo{year}{2020}\natexlab{a}.
\newblock \showarticletitle{Inquisitive question generation for high level text comprehension}.
\newblock \bibinfo{journal}{\emph{arXiv preprint arXiv:2010.01657}} (\bibinfo{year}{2020}).
\newblock


\bibitem[Ko et~al\mbox{.}(2020b)]%
        {inquisitive-online}
\bibfield{author}{\bibinfo{person}{Wei-Jen Ko}, \bibinfo{person}{Te-yuan Chen}, \bibinfo{person}{Yiyan Huang}, \bibinfo{person}{Greg Durrett}, {and} \bibinfo{person}{Junyi~Jessy Li}.} \bibinfo{year}{2020}\natexlab{b}.
\newblock \bibinfo{booktitle}{\emph{Inquitive Dataset}}.
\newblock
\urldef\tempurl%
\url{https://github.com/wjko2/INQUISITIVE}
\showURL{%
\tempurl}


\bibitem[Kong et~al\mbox{.}(2015)]%
        {kong2015predicting}
\bibfield{author}{\bibinfo{person}{Weize Kong}, \bibinfo{person}{Rui Li}, \bibinfo{person}{Jie Luo}, \bibinfo{person}{Aston Zhang}, \bibinfo{person}{Yi Chang}, {and} \bibinfo{person}{James Allan}.} \bibinfo{year}{2015}\natexlab{}.
\newblock \showarticletitle{Predicting search intent based on pre-search context}. In \bibinfo{booktitle}{\emph{Proceedings of the 38th International ACM SIGIR Conference on Research and Development in Information Retrieval}}. \bibinfo{pages}{503--512}.
\newblock


\bibitem[Koskela et~al\mbox{.}(2018)]%
        {koskela2018proactive}
\bibfield{author}{\bibinfo{person}{Markus Koskela}, \bibinfo{person}{Petri Luukkonen}, \bibinfo{person}{Tuukka Ruotsalo}, \bibinfo{person}{Mats Sj{\"o}berg}, {and} \bibinfo{person}{Patrik Flor{\'e}en}.} \bibinfo{year}{2018}\natexlab{}.
\newblock \showarticletitle{Proactive information retrieval by capturing search intent from primary task context}.
\newblock \bibinfo{journal}{\emph{ACM Transactions on Interactive Intelligent Systems (TiiS)}} \bibinfo{volume}{8}, \bibinfo{number}{3} (\bibinfo{year}{2018}), \bibinfo{pages}{1--25}.
\newblock


\bibitem[Lee et~al\mbox{.}(2018)]%
        {lee2018making}
\bibfield{author}{\bibinfo{person}{Sunhwan Lee}, \bibinfo{person}{Robert Moore}, \bibinfo{person}{Guang-Jie Ren}, \bibinfo{person}{Raphael Arar}, {and} \bibinfo{person}{Shun Jiang}.} \bibinfo{year}{2018}\natexlab{}.
\newblock \showarticletitle{Making personalized recommendation through conversation: Architecture design and recommendation methods}. In \bibinfo{booktitle}{\emph{Workshops at the Thirty-Second AAAI Conference on Artificial Intelligence}}.
\newblock


\bibitem[Lewis et~al\mbox{.}(2020)]%
        {lewis2020retrieval}
\bibfield{author}{\bibinfo{person}{Patrick Lewis}, \bibinfo{person}{Ethan Perez}, \bibinfo{person}{Aleksandra Piktus}, \bibinfo{person}{Fabio Petroni}, \bibinfo{person}{Vladimir Karpukhin}, \bibinfo{person}{Naman Goyal}, \bibinfo{person}{Heinrich K{\"u}ttler}, \bibinfo{person}{Mike Lewis}, \bibinfo{person}{Wen-tau Yih}, \bibinfo{person}{Tim Rockt{\"a}schel}, {et~al\mbox{.}}} \bibinfo{year}{2020}\natexlab{}.
\newblock \showarticletitle{Retrieval-augmented generation for knowledge-intensive nlp tasks}.
\newblock \bibinfo{journal}{\emph{Advances in Neural Information Processing Systems}}  \bibinfo{volume}{33} (\bibinfo{year}{2020}), \bibinfo{pages}{9459--9474}.
\newblock


\bibitem[Li et~al\mbox{.}(2014)]%
        {li2014semantic}
\bibfield{author}{\bibinfo{person}{Hang Li}, \bibinfo{person}{Jun Xu}, {et~al\mbox{.}}} \bibinfo{year}{2014}\natexlab{}.
\newblock \showarticletitle{Semantic matching in search}.
\newblock \bibinfo{journal}{\emph{Foundations and Trends{\textregistered} in Information Retrieval}} \bibinfo{volume}{7}, \bibinfo{number}{5} (\bibinfo{year}{2014}), \bibinfo{pages}{343--469}.
\newblock


\bibitem[Liebling et~al\mbox{.}(2012)]%
        {liebling2012anticipatory}
\bibfield{author}{\bibinfo{person}{Daniel~J Liebling}, \bibinfo{person}{Paul~N Bennett}, {and} \bibinfo{person}{Ryen~W White}.} \bibinfo{year}{2012}\natexlab{}.
\newblock \showarticletitle{Anticipatory search: using context to initiate search}. In \bibinfo{booktitle}{\emph{Proceedings of the 35th international ACM SIGIR conference on Research and development in information retrieval}}. \bibinfo{pages}{1035--1036}.
\newblock


\bibitem[Lin et~al\mbox{.}(2021)]%
        {lin2021pyserini}
\bibfield{author}{\bibinfo{person}{Jimmy Lin}, \bibinfo{person}{Xueguang Ma}, \bibinfo{person}{Sheng-Chieh Lin}, \bibinfo{person}{Jheng-Hong Yang}, \bibinfo{person}{Ronak Pradeep}, {and} \bibinfo{person}{Rodrigo Nogueira}.} \bibinfo{year}{2021}\natexlab{}.
\newblock \showarticletitle{Pyserini: A Python toolkit for reproducible information retrieval research with sparse and dense representations}. In \bibinfo{booktitle}{\emph{Proceedings of the 44th International ACM SIGIR Conference on Research and Development in Information Retrieval}}. \bibinfo{pages}{2356--2362}.
\newblock


\bibitem[Lin et~al\mbox{.}(2022)]%
        {lin2022pretrained}
\bibfield{author}{\bibinfo{person}{Jimmy Lin}, \bibinfo{person}{Rodrigo Nogueira}, {and} \bibinfo{person}{Andrew Yates}.} \bibinfo{year}{2022}\natexlab{}.
\newblock \bibinfo{booktitle}{\emph{Pretrained transformers for text ranking: Bert and beyond}}.
\newblock \bibinfo{publisher}{Springer Nature}.
\newblock


\bibitem[Liu et~al\mbox{.}(2023)]%
        {liu2023pre}
\bibfield{author}{\bibinfo{person}{Pengfei Liu}, \bibinfo{person}{Weizhe Yuan}, \bibinfo{person}{Jinlan Fu}, \bibinfo{person}{Zhengbao Jiang}, \bibinfo{person}{Hiroaki Hayashi}, {and} \bibinfo{person}{Graham Neubig}.} \bibinfo{year}{2023}\natexlab{}.
\newblock \showarticletitle{Pre-train, prompt, and predict: A systematic survey of prompting methods in natural language processing}.
\newblock \bibinfo{journal}{\emph{Comput. Surveys}} \bibinfo{volume}{55}, \bibinfo{number}{9} (\bibinfo{year}{2023}), \bibinfo{pages}{1--35}.
\newblock


\bibitem[MERROUNI et~al\mbox{.}(2019)]%
        {merrouni2019toward}
\bibfield{author}{\bibinfo{person}{Zakariae~ALAMI MERROUNI}, \bibinfo{person}{Bouchra FRIKH}, {and} \bibinfo{person}{Brahim OUHBI}.} \bibinfo{year}{2019}\natexlab{}.
\newblock \showarticletitle{Toward contextual information retrieval: a review and trends}.
\newblock \bibinfo{journal}{\emph{Procedia computer science}}  \bibinfo{volume}{148} (\bibinfo{year}{2019}), \bibinfo{pages}{191--200}.
\newblock


\bibitem[Microsoft(2024)]%
        {ms-marco-website}
\bibfield{author}{\bibinfo{person}{Microsoft}.} \bibinfo{year}{2024}\natexlab{}.
\newblock \bibinfo{booktitle}{\emph{MS MARCO}}.
\newblock
\urldef\tempurl%
\url{https://microsoft.github.io/msmarco/}
\showURL{%
\tempurl}


\bibitem[Nguyen et~al\mbox{.}(2016)]%
        {nguyen2016ms}
\bibfield{author}{\bibinfo{person}{Tri Nguyen}, \bibinfo{person}{Mir Rosenberg}, \bibinfo{person}{Xia Song}, \bibinfo{person}{Jianfeng Gao}, \bibinfo{person}{Saurabh Tiwary}, \bibinfo{person}{Rangan Majumder}, {and} \bibinfo{person}{Li Deng}.} \bibinfo{year}{2016}\natexlab{}.
\newblock \showarticletitle{MS MARCO: A human generated machine reading comprehension dataset}.
\newblock \bibinfo{journal}{\emph{choice}}  \bibinfo{volume}{2640} (\bibinfo{year}{2016}), \bibinfo{pages}{660}.
\newblock


\bibitem[PyTorch(2024)]%
        {cosineembloss}
\bibfield{author}{\bibinfo{person}{PyTorch}.} \bibinfo{year}{2024}\natexlab{}.
\newblock \bibinfo{booktitle}{\emph{CosineEmbeddingLoss - PyTorch 2.3 Documentation}}.
\newblock
\urldef\tempurl%
\url{https://huggingface.co/google-bert/bert-base-uncased}
\showURL{%
\tempurl}


\bibitem[Radford et~al\mbox{.}(2019)]%
        {radford2019language}
\bibfield{author}{\bibinfo{person}{Alec Radford}, \bibinfo{person}{Jeffrey Wu}, \bibinfo{person}{Rewon Child}, \bibinfo{person}{David Luan}, \bibinfo{person}{Dario Amodei}, \bibinfo{person}{Ilya Sutskever}, {et~al\mbox{.}}} \bibinfo{year}{2019}\natexlab{}.
\newblock \showarticletitle{Language models are unsupervised multitask learners}.
\newblock \bibinfo{journal}{\emph{OpenAI blog}} \bibinfo{volume}{1}, \bibinfo{number}{8} (\bibinfo{year}{2019}), \bibinfo{pages}{9}.
\newblock


\bibitem[Raffel et~al\mbox{.}(2020)]%
        {raffel2020exploring}
\bibfield{author}{\bibinfo{person}{Colin Raffel}, \bibinfo{person}{Noam Shazeer}, \bibinfo{person}{Adam Roberts}, \bibinfo{person}{Katherine Lee}, \bibinfo{person}{Sharan Narang}, \bibinfo{person}{Michael Matena}, \bibinfo{person}{Yanqi Zhou}, \bibinfo{person}{Wei Li}, {and} \bibinfo{person}{Peter~J Liu}.} \bibinfo{year}{2020}\natexlab{}.
\newblock \showarticletitle{Exploring the limits of transfer learning with a unified text-to-text transformer}.
\newblock \bibinfo{journal}{\emph{The Journal of Machine Learning Research}} \bibinfo{volume}{21}, \bibinfo{number}{1} (\bibinfo{year}{2020}), \bibinfo{pages}{5485--5551}.
\newblock


\bibitem[Rahurkar and Cucerzan(2008)]%
        {rahurkar2008predicting}
\bibfield{author}{\bibinfo{person}{Mandar Rahurkar} {and} \bibinfo{person}{Silviu Cucerzan}.} \bibinfo{year}{2008}\natexlab{}.
\newblock \showarticletitle{Predicting when browsing context is relevant to search}. In \bibinfo{booktitle}{\emph{Proceedings of the 31st Annual international ACM SIGIR Conference on Research and Development in information Retrieval}}. \bibinfo{pages}{841--842}.
\newblock


\bibitem[Reddy et~al\mbox{.}(2019)]%
        {reddy2019coqa}
\bibfield{author}{\bibinfo{person}{Siva Reddy}, \bibinfo{person}{Danqi Chen}, {and} \bibinfo{person}{Christopher~D Manning}.} \bibinfo{year}{2019}\natexlab{}.
\newblock \showarticletitle{Coqa: A conversational question answering challenge}.
\newblock \bibinfo{journal}{\emph{Transactions of the Association for Computational Linguistics}}  \bibinfo{volume}{7} (\bibinfo{year}{2019}), \bibinfo{pages}{249--266}.
\newblock


\bibitem[Reimers and Gurevych(2019)]%
        {reimers-2019-sentence-bert}
\bibfield{author}{\bibinfo{person}{Nils Reimers} {and} \bibinfo{person}{Iryna Gurevych}.} \bibinfo{year}{2019}\natexlab{}.
\newblock \showarticletitle{Sentence-BERT: Sentence Embeddings using Siamese BERT-Networks}. In \bibinfo{booktitle}{\emph{Proceedings of the 2019 Conference on Empirical Methods in Natural Language Processing}}. \bibinfo{publisher}{Association for Computational Linguistics}.
\newblock
\urldef\tempurl%
\url{https://arxiv.org/abs/1908.10084}
\showURL{%
\tempurl}


\bibitem[Rhodes and Starner(1996)]%
        {rhodes1996remembrance}
\bibfield{author}{\bibinfo{person}{Bradley Rhodes} {and} \bibinfo{person}{Thad Starner}.} \bibinfo{year}{1996}\natexlab{}.
\newblock \showarticletitle{Remembrance Agent: A continuously running automated information retrieval system}. In \bibinfo{booktitle}{\emph{The proceedings of the first international conference on the practical application of intelligent agents and multi agent technology}}, Vol.~\bibinfo{volume}{1}. \bibinfo{pages}{487--495}.
\newblock


\bibitem[Robertson and Walker(1994)]%
        {robertson1994some}
\bibfield{author}{\bibinfo{person}{Stephen~E Robertson} {and} \bibinfo{person}{Steve Walker}.} \bibinfo{year}{1994}\natexlab{}.
\newblock \showarticletitle{Some simple effective approximations to the 2-poisson model for probabilistic weighted retrieval}. In \bibinfo{booktitle}{\emph{SIGIR’94: Proceedings of the Seventeenth Annual International ACM-SIGIR Conference on Research and Development in Information Retrieval, organised by Dublin City University}}. Springer, \bibinfo{pages}{232--241}.
\newblock


\bibitem[Smucker et~al\mbox{.}(2007)]%
        {smucker2007comparison}
\bibfield{author}{\bibinfo{person}{Mark~D Smucker}, \bibinfo{person}{James Allan}, {and} \bibinfo{person}{Ben Carterette}.} \bibinfo{year}{2007}\natexlab{}.
\newblock \showarticletitle{A comparison of statistical significance tests for information retrieval evaluation}. In \bibinfo{booktitle}{\emph{Proceedings of the sixteenth ACM conference on Conference on information and knowledge management}}. \bibinfo{pages}{623--632}.
\newblock


\bibitem[Song and Guo(2016)]%
        {song2016query}
\bibfield{author}{\bibinfo{person}{Yang Song} {and} \bibinfo{person}{Qi Guo}.} \bibinfo{year}{2016}\natexlab{}.
\newblock \showarticletitle{Query-less: Predicting task repetition for nextgen proactive search and recommendation engines}. In \bibinfo{booktitle}{\emph{Proceedings of the 25th International Conference on World Wide Web}}. \bibinfo{pages}{543--553}.
\newblock


\bibitem[Tahery and Farzi(2020)]%
        {tahery2020customized}
\bibfield{author}{\bibinfo{person}{Saedeh Tahery} {and} \bibinfo{person}{Saeed Farzi}.} \bibinfo{year}{2020}\natexlab{}.
\newblock \showarticletitle{Customized query auto-completion and suggestion—A review}.
\newblock \bibinfo{journal}{\emph{Information Systems}}  \bibinfo{volume}{87} (\bibinfo{year}{2020}), \bibinfo{pages}{101415}.
\newblock


\bibitem[Taylor(1962)]%
        {taylor1962process}
\bibfield{author}{\bibinfo{person}{Robert~S Taylor}.} \bibinfo{year}{1962}\natexlab{}.
\newblock \showarticletitle{The process of asking questions}.
\newblock \bibinfo{journal}{\emph{American documentation}} \bibinfo{volume}{13}, \bibinfo{number}{4} (\bibinfo{year}{1962}), \bibinfo{pages}{391--396}.
\newblock


\bibitem[Torbati et~al\mbox{.}(2021)]%
        {torbati2021you}
\bibfield{author}{\bibinfo{person}{Ghazaleh~H Torbati}, \bibinfo{person}{Andrew Yates}, {and} \bibinfo{person}{Gerhard Weikum}.} \bibinfo{year}{2021}\natexlab{}.
\newblock \showarticletitle{You get what you chat: Using conversations to personalize search-based recommendations}. In \bibinfo{booktitle}{\emph{Advances in Information Retrieval: 43rd European Conference on IR Research, ECIR 2021, Virtual Event, March 28--April 1, 2021, Proceedings, Part I 43}}. Springer, \bibinfo{pages}{207--223}.
\newblock


\bibitem[Touvron et~al\mbox{.}(2023)]%
        {touvron2023llama}
\bibfield{author}{\bibinfo{person}{Hugo Touvron}, \bibinfo{person}{Louis Martin}, \bibinfo{person}{Kevin Stone}, \bibinfo{person}{Peter Albert}, \bibinfo{person}{Amjad Almahairi}, \bibinfo{person}{Yasmine Babaei}, \bibinfo{person}{Nikolay Bashlykov}, \bibinfo{person}{Soumya Batra}, \bibinfo{person}{Prajjwal Bhargava}, \bibinfo{person}{Shruti Bhosale}, {et~al\mbox{.}}} \bibinfo{year}{2023}\natexlab{}.
\newblock \showarticletitle{Llama 2: Open foundation and fine-tuned chat models}.
\newblock \bibinfo{journal}{\emph{arXiv preprint arXiv:2307.09288}} (\bibinfo{year}{2023}).
\newblock


\bibitem[Van~Gysel(2017)]%
        {van2017remedies}
\bibfield{author}{\bibinfo{person}{Christophe Van~Gysel}.} \bibinfo{year}{2017}\natexlab{}.
\newblock \showarticletitle{Remedies against the vocabulary gap in information retrieval}.
\newblock \bibinfo{journal}{\emph{arXiv preprint arXiv:1711.06004}} (\bibinfo{year}{2017}).
\newblock


\bibitem[Vaswani et~al\mbox{.}(2017)]%
        {vaswani2017attention}
\bibfield{author}{\bibinfo{person}{Ashish Vaswani}, \bibinfo{person}{Noam Shazeer}, \bibinfo{person}{Niki Parmar}, \bibinfo{person}{Jakob Uszkoreit}, \bibinfo{person}{Llion Jones}, \bibinfo{person}{Aidan~N Gomez}, \bibinfo{person}{{\L}ukasz Kaiser}, {and} \bibinfo{person}{Illia Polosukhin}.} \bibinfo{year}{2017}\natexlab{}.
\newblock \showarticletitle{Attention is all you need}.
\newblock \bibinfo{journal}{\emph{Advances in neural information processing systems}}  \bibinfo{volume}{30} (\bibinfo{year}{2017}).
\newblock


\bibitem[Virtanen et~al\mbox{.}(2020)]%
        {2020SciPy-NMeth}
\bibfield{author}{\bibinfo{person}{Pauli Virtanen}, \bibinfo{person}{Ralf Gommers}, \bibinfo{person}{Travis~E. Oliphant}, \bibinfo{person}{Matt Haberland}, \bibinfo{person}{Tyler Reddy}, \bibinfo{person}{David Cournapeau}, \bibinfo{person}{Evgeni Burovski}, \bibinfo{person}{Pearu Peterson}, \bibinfo{person}{Warren Weckesser}, \bibinfo{person}{Jonathan Bright}, \bibinfo{person}{St{\'e}fan~J. {van der Walt}}, \bibinfo{person}{Matthew Brett}, \bibinfo{person}{Joshua Wilson}, \bibinfo{person}{K.~Jarrod Millman}, \bibinfo{person}{Nikolay Mayorov}, \bibinfo{person}{Andrew R.~J. Nelson}, \bibinfo{person}{Eric Jones}, \bibinfo{person}{Robert Kern}, \bibinfo{person}{Eric Larson}, \bibinfo{person}{C~J Carey}, \bibinfo{person}{{\.I}lhan Polat}, \bibinfo{person}{Yu Feng}, \bibinfo{person}{Eric~W. Moore}, \bibinfo{person}{Jake {VanderPlas}}, \bibinfo{person}{Denis Laxalde}, \bibinfo{person}{Josef Perktold}, \bibinfo{person}{Robert Cimrman}, \bibinfo{person}{Ian Henriksen}, \bibinfo{person}{E.~A. Quintero},
  \bibinfo{person}{Charles~R. Harris}, \bibinfo{person}{Anne~M. Archibald}, \bibinfo{person}{Ant{\^o}nio~H. Ribeiro}, \bibinfo{person}{Fabian Pedregosa}, \bibinfo{person}{Paul {van Mulbregt}}, {and} \bibinfo{person}{{SciPy 1.0 Contributors}}.} \bibinfo{year}{2020}\natexlab{}.
\newblock \showarticletitle{{{SciPy} 1.0: Fundamental Algorithms for Scientific Computing in Python}}.
\newblock \bibinfo{journal}{\emph{Nature Methods}}  \bibinfo{volume}{17} (\bibinfo{year}{2020}), \bibinfo{pages}{261--272}.
\newblock
\urldef\tempurl%
\url{https://doi.org/10.1038/s41592-019-0686-2}
\showDOI{\tempurl}


\bibitem[White et~al\mbox{.}(2010)]%
        {white2010predicting}
\bibfield{author}{\bibinfo{person}{Ryen~W White}, \bibinfo{person}{Paul~N Bennett}, {and} \bibinfo{person}{Susan~T Dumais}.} \bibinfo{year}{2010}\natexlab{}.
\newblock \showarticletitle{Predicting short-term interests using activity-based search context}. In \bibinfo{booktitle}{\emph{Proceedings of the 19th ACM international conference on Information and knowledge management}}. \bibinfo{pages}{1009--1018}.
\newblock


\end{thebibliography}
\bibliographystyle{ACM-Reference-Format}

\begin{comment}
\appendix
\section{Prompt for Splitting Queries}\label{split query prompt}
...put prompt here
\section{Dataset Samples}\label{dataset samples}
...put dataset samples here

\subsection{Encoder-Decoder}\label{encoder decoder samples}
example input and hyperparameters

\subsection{LLM}\label{gen llm samples}
example input

\subsection{Bi-Encoder}\label{biencoder samples}
example input and hyperparameters

\subsection{Cross-Encoder}\label{crossencoder samples}
example input and hyperparameters
\end{comment}

\end{document}